\documentstyle [prl,aps]{revtex}
\input epsf
\begin{document}
\draft
\title{First Dark Matter Limits from a Large-Mass, Low-Background\\ 
Superheated Droplet Detector}
\author{
J.I. Collar$^{a,b,*}$, J. Puibasset $^{a}$, T.A. Girard$^{c}$, D. Limagne$^{a}$,
 H.S. Miley$^{d}$ and G. Waysand$^{a}$
}
\address{ 
$^{a}$Groupe de Physique des Solides (UMR CNRS 75-88), Universit\'es Paris 
7 \& 6, 
75251 Paris Cedex 05, France\\
$^{b}$CERN, EP Division, CH-1211 Geneve 23, Switzerland\\
$^{c}$Centro de F\'\i sica Nuclear, Universidade de Lisboa, 1649-003 Lisbon, 
Portugal\\
$^{d}$Pacific Northwest National Laboratory, Richland, WA 99352, USA
}
\wideabs{
\maketitle
\begin{abstract}
\widetext
We report on the fabrication aspects and calibration of 
the first large active mass ($\sim\!15$ g) modules of SIMPLE, 
a search 
for particle dark 
matter using Superheated Droplet Detectors (SDDs). 
While still 
limited by the statistical uncertainty of the small 
data sample on hand, the 
first weeks of operation in the new underground laboratory 
of Rustrel-Pays 
d'Apt already provide a sensitivity to axially-coupled 
Weakly Interacting 
Massive Particles (WIMPs) competitive with leading 
experiments, confirming  
SDDs as a convenient, low-cost alternative for WIMP 
detection. \end{abstract}
\pacs{PACS number(s): 
95.35.+d, 29.40.-n, 05.70.Fh\\
$^{*}$~Corresponding author. E-mail: 
Juan.Collar@cern.ch}} \narrowtext
The rupture of metastability by radiation has been 
historically exploited as a method for particle detection. 
Perhaps its most successful application is the Bubble 
Chamber, 
where ionizing particles deposit enough local energy in a 
superheated liquid to produce vaporization along their wake. Apfel 
extended this concept in the form of Superheated Droplet Detectors 
\cite{apfel1} (SDDs, a.k.a. Bubble Detectors), 
in which small drops (radius 
$\sim10~\mu m$) 
of the liquid are uniformly dispersed in a gel or viscoelastic 
medium. In a SDD the gel matrix 
isolates the fragile metastable system from vibrations and 
convection currents, while the smooth liquid-liquid interfaces 
impede the continuous triggering on surface impurities
that occurs 
in bubble chambers. The lifetime of the 
superheated state is
extended, allowing for new applications: 
SDDs are increasingly 
popular as neutron dosimeters, where the nucleated 
visible bubbles provide a reading of the radiation exposure.
SIMPLE ({\underline S}uperheated {\underline I}nstrument 
for {\underline
M}assive {\underline P}artic{\underline {LE}} searches) aims to 
detect particle dark matter using SDDs. We report here on the 
sensitivity attained at the early prototype stage, already comparable
to the best achieved with competing technologies.

In the moderately superheated industrial refrigerants used 
in SDDs, bubbles are produced only by particles having 
elevated stopping powers ($dE/dx\gtrsim 200 ~keV/\mu m$) 
such as nuclear recoils. This is understood in the 
framework of the ``thermal spike'' model \cite{seitz}, 
common to bubble chambers: for the transition to occur, a 
vapor 
nucleus of radius $> r_{c}$ must be created, while only the 
energy 
deposited along a distance comparable to this critical radius 
$r_{c}$ is available for its formation. Hence, a double 
threshold is imposed: the deposited energy  
$E$
must be larger than the
work of formation of the critical nucleus, $E_{c}$, and this 
energy must be lost 
over a distance $\textrm{O}\!\left(r_{c}\right)$, 
i.e., a minimum $dE/dx$ 
is required. More formally \cite{apfel2,peyrou}: 
\begin{eqnarray}
E>E_{c}=4\pi r^{2}_{c} \gamma/3\epsilon \nonumber\\
dE/dx>E_{c}/a r_{c},
\end{eqnarray}
where $r_{c}\!=\!2~\gamma/\Delta P$, 
$\gamma\!\left(T\right)$ is the surface 
tension, $\Delta P\!=\!P_{V}\!-\!P$, 
$P_{V}\!\left(T\right) $ 
is the vapor pressure, $P$ and $T$ are the operating pressure and 
temperature, $\epsilon$ varies in the range 
$[0.02,0.06]$ for different liquids \cite{peyrou,apfelroy}, 
and $a\!\left(T\right)\!\sim \! \textrm{O}\!\left(1\right)$ \cite{harper}. 

Both thresholds can be tuned by changing 
the operating conditions: keV nuclear 
recoils like those expected from scattering of WIMPs 
(currently the favored galactic dark matter candidates \cite{wimps}) 
are detectable at room $T$ and atmospheric $P$, 
allowing for a low-cost search free of 
the complications associated to cryogenic equipment. 
Most importantly, the threshold in
$dE/dx$  provides an insensitivity to
minimum-ionizing backgrounds 
that hamper the numerous WIMP detection efforts \cite{review}.
A mere $<\!\!10$ WIMP recoils/kg target/day 
are expected and hence the 
importance of background reduction and/or rejection. 
SDDs of active mass O(1)kg can in principle considerably
extend the present experimental 
sensitivity \cite{myprd}. 

Prompted by the modest active mass of commercially available 
SDDs 
($\sim\!0.03$ g refrigerant/dosimeter) and the need to control the 
fabrication process, we developed
a $80~l$, 60 bar pressure reactor dedicated to  
large-mass SDD production. It
houses a variable-speed magnetic stirrer, 
heating and cooling elements and micropumps 
for catalyst addition (we nevertheless favored thermally-reversible 
food gels due to safety 
concerns in the handling of synthetic monomers). The 
fabrication of $1~l$ SDD 
modules containing up to 3\% in superheated liquid starts 
with the preparation of a suitable gel matrix;
ingredients are 
selected and processed in order to avoid alpha emitters, 
the only internal radioemitters of concern
\cite{myprd}. A precise density matching between matrix 
and refrigerant is needed to obtain a uniform droplet 
dispersion, making 
water-based gels inadequate unless large fractions of 
inorganic 
salts are added, which can unbalance the chemistry of the 
composite and contribute an undesirable concentration of 
these contaminants \cite{zacek2}. We find that glycerol is 
for this and other
reasons an additive of choice.
It is purified using a bed of pre-eluted ion-exchanging resin 
specifically targeted at 
actinide removal. Polymer additives and gelating agent  
are washed in a resin bath. All components are forced through 
0.2 $\mu m$ filters to remove motes that 
can act as nucleation centers. The resulting mixture is outgassed 
and maintained above its gelation temperature in the reactor. 
The refrigerant is distilled
and incorporated to this solution at
$P>>P_{V}\!\left(T\right)$ to avoid boiling during the ensuing 
vigorous stirring. After a homogenized dispersion 
($r\sim30\pm15~\mu$m) of droplets is 
obtained, cooling, setting and step-wise adiabatic 
decompression produce 
a delicate entanglement of superheated liquid 
and thermally-reversible gel, the SDD.  
The detectors are refrigerated and pressurized 
during storage to inhibit their response to 
environmental 
neutrons. 

SDDs can bypass the listed problems 
associated to a former \cite{zacek} bubble chamber WIMP search 
proposal, but are not devoid of their own idiosyncrasies. 
For instance, the solubility 
of hydrogen-free refrigerant liquids in water-based gels is small 
(e.g., 0.002 mol/kg bar for R-12, $CCl_{2}F_{2}$),  yet 
sufficient to produce unchecked bubble growth via permeation
after few days of continuous SDD operation. 
The engorged bubbles lead to fractures, spurious 
nucleations and depletion of the superheated liquid 
(commercial gel-based SDDs are 
designed for few hours of exposure before recompression 
\cite{apfelco}, 
a cycle that can be repeated a limited number of times).
To achieve the long-term SDD stability needed for a WIMP search, we 
employ a multiple strategy: fracture formation 
can be delayed 
under a moderate $P\!\sim\! 2$ atm, or by choosing 
refrigerants with the lowest solubility in the 
matrix ($\sim 0.0003$ 
mol/kg bar for R-115, $C_{2}ClF_{5}$). Structure-making 
inorganic salts produce a "salting-out" effect, 
i.e., further 
reduce the refrigerant solubility. Their use being inadvisable 
for the reasons above, we introduce instead 
polymers known to have a similar effect \cite{pvp}, such as 
polyvinylpyrrolidone (PVP). As a result of these 
measures, present SIMPLE modules are stable over 
$\sim$40 d of continuous exposure. Another example of SDD-specific 
problems is the formation of 
clathrate-hydrates
on droplet boundaries during fabrication or recompression. These 
metastable ice-like structures are
inclusions of refrigerant molecules into water 
cages \cite{cages} that shorten the 
lifetime of  superheated drops encrusted by 
them via transfer 
mechanisms still not well understood \cite{mori}. Their presence 
may be responsible for a long-lived spurious 
nucleation rate observed 
in R-12 SDDs following fabrication \cite{roy}. This 
is addressed 
in SIMPLE with the addition 
of polymers such as PVCap or PVP, 
which act as kinetic 
inhibitors in their growth \cite{cages}, and by 
use of large molecular size refrigerants like R-115, 
for which the formation of most hydrates is 
stoichiometrically forbidden \cite{cages,mori}. 

Prototype modules are tested in an underground
gallery. The 27 m rock
overburden and $\sim\! 30$ cm paraffin shielding 
reduce the flux of muon-induced and cosmic fast 
neutrons,
the main source of 
nucleations above ground. 
Inside the shielding, a water$+$glycol thermally-regulated 
bath maintains $T$ constant to within 
$0.1^{\circ}$C. 
The characteristic violent sound pulse accompanying 
vaporization in superheated liquids \cite{sound1,sound2}
is picked-up by a small piezoelectric transducer in the 
interior of the module, amplified 
and stored. Special precautions are 
taken against acoustic and 
seismic noise. \frenchspacing{Fig. 1} displays 
the decrease 
in spontaneous bubble 
nucleation rate brought by 
progressive purification of the 
modules.
\begin{figure}[tbp]
\epsfxsize = \hsize 
\epsfbox{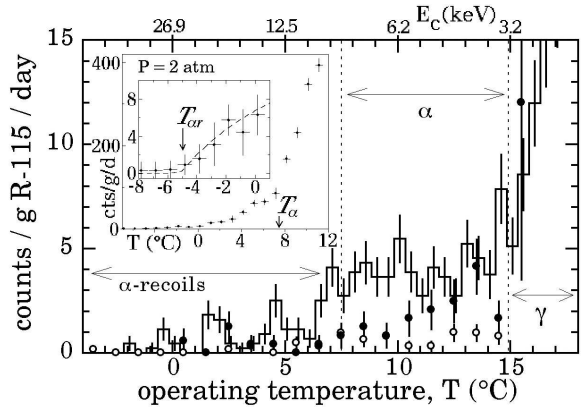} 
\caption{SDD background at 90 
m.w.e. and $P\!\!=$2 atm,
following cumulative steps of cleansing; histogram: double distillation of 
water and microfiltration, $\bullet$: distillation 
of refrigerant and glycerin purification, $\circ$: gelatine and 
PVP purification. {\it Insert:} Rate induced in a
calibration (0.2
Bq $^{241}$Am/g gel). The {\it theoretical} values of $T_{\alpha r}$ and 
$T_{\alpha}$  are indicated (the dashed line 
is an numerical simulation containing no 
free parameters).} 
\end{figure}

The response of smaller SDDs to various 
neutron fields has been extensively studied \cite{harper,apfel3}
and found to match theoretical 
expectations. However, large-size, opaque SDDs 
require independent calibration: 
acoustic detection of the explosion of the smallest or most distant 
droplets is not {\it a priori} guaranteed. The  
energy released as sound varies as 
$(P_{V}-P)^{3/2}$ \cite{sound2}, making these additional 
characterizations even more imperative 
for SDDs operated under $P\!>\!1$ atm. Two separate 
types of calibration have 
been performed to determine the target mass  effectively monitored
in SIMPLE modules and to check the calculation of the $T,P$-dependent
threshold energy $E_{thr}$ above which WIMP recoils can induce
nucleations (defined as the lowest energy meeting both 
conditions in \frenchspacing{Eq. (1)} \cite{myprd,nagdy,derico}). 
First, a liquid $^{241}$Am source
(an alpha emitter) is diluted into the matrix while still in the 
solution state. 
Following Eq. (1), the 5.5 MeV alphas and 91 keV recoiling $^{237}$Np 
daughters cannot induce nucleations at temperatures below $T_{\alpha}$ 
and 
$T_{\alpha r}$, respectively \cite{myprd}.
The expression  $a=4.3 
\left(\rho_{v}/\rho_{l}\right)^{1/3}$ \cite{harper}, 
where $\rho_{v}(T)$, $\rho_{l}(T)$ are the vapor- and liquid-phase 
densities of the refrigerant, correctly predicts  
the observed $T_{\alpha}$ for both R-12 
and 
R-115 at $P\!=$1 and 2 atm. In the same 
conditions, the theoretical value of 
$\epsilon$ 
\cite{note} for 
these liquids ($\epsilon\sim\!0.026$, neglecting a small $T,P$ 
dependence) generates a good agreement with the experimental 
$T_{\alpha r}$ (Fig. 
1, insert).
Prior to extensive component purification, 
the spectrum in non-calibration runs (\frenchspacing{Fig. 1}, histogram) 
bears close resemblance to that produced by $^{241}$Am 
spiking (\frenchspacing{Fig. 1}, insert); 
the initial presence of a small ($\sim\! 10^{-4}$ pCi/g) $^{228}$Th 
contamination, compatible with the observed rate, was confirmed 
via low-level alpha spectroscopy. Three regimes of background 
dominance are therefore delimited by vertical lines in \frenchspacing{Fig. 
1}: the sudden rise 
at $T\!\sim\!15^{\circ}$C originates 
in high-$dE/dx$ Auger electron cascades 
following interactions of environmental gammas with Cl atoms in the 
refrigerant \cite{peyrou,hahn}. The calculated $E_{c}$ for R-115 at 
$T\!=\!15.5^{\circ}$C and 
$\!P\!=$2 atm is 2.9 keV, coincidental with the binding energy of 
K-shell electrons in Cl, 2.8 keV (i.e., the maximum $E$ deposited 
via this mechanism). Thus, the onset of gamma sensitivity provides 
a welcome additional check of the threshold in the few keV region. 

Alpha calibrations are not suitable for a rigorous determination of 
the overall sound detection efficiency because a large fraction of 
the added emitters drifts to gel-droplet boundaries 
during fabrication, an effect explained by the polarity of 
actinide complex ions \cite{wang} and dependent on matrix 
composition. 
While this migration does not affect $T_{\alpha}$ nor $T_{\alpha 
r}$, it enhances the overall nucleation efficiency in a somewhat 
unpredictable manner \cite{wang}. To make up for this deficiency, 
SIMPLE modules have been exposed to a 
$^{252}$Cf neutron 
source at the TIS/RP calibration facility (CERN). The 
resulting spectrum of neutron-induced fluorine recoils 
(\frenchspacing{Fig. 2}, insert) mimics 
a typically expected one from WIMP interactions. A complete 
MCNP4a \cite{mcnp} simulation of the calibration setup takes 
into account the contribution from albedo and thermal neutrons.
The expected nucleation rate as a function of $T$ is calculated 
as in \cite{myprd,apfel3}: cross sections for the elastic, 
inelastic, (n,$\alpha$) and (n,p) channels of the refrigerant 
constituents 
are extracted from ENDFB-VI libraries. Look-up 
tables of the distribution of deposited energies as a 
function of neutron energy are built from the SPECTER code 
\cite{specter}, stopping powers of the recoiling species 
are taken from SRIM98 \cite{trim}.
Since $T$ was continuously ramped up during the 
irradiations at 
a relatively fast 1.1$^{\circ}$C/hr, 
a small correction to it ($<\!1^{\circ}$C) is 
numerically computed and applied to account for 
the slow thermalization of the module.
Depending on $T$, the value 
of $E_{thr}$ for elastic recoils in fluorine (the dominant 
nucleation mechanism in R-115) is set by either condition in 
\frenchspacing{Eq. (1)}, the other being always fulfilled for 
$E>E_{thr}$ \cite{myprd,nagdy}. The handover from the second 
to the
first condition at $T$ above $\sim 5.5^{\circ}$C ($\sim 2.5^{\circ}$C) 
for $P\!=$2 atm ($P\!=$1 atm) 
is clearly observed in the data 
as two different regimes of nucleation rate (\frenchspacing{Fig. 2}). 
A larger-than-expected response, already noticed 
in R-12 \cite{harper}, is evident at low $T$: the calculated $E_{thr}$ 
there is too conservative (too high). This behavior appears well 
below the normal regime of SDD operation (which is at $T$ high 
enough to have $E_{thr}\!=\!E_{c}$) and therefore does not interfere 
with neutron or WIMP detection. However, it is interesting in that it 
points at a higher than normal bubble nucleation 
efficiency from heavy particles, 
as discussed in early bubble chamber work \cite{hahn}.
A best-fit to the overall normalization of the Monte Carlo over the 
full data set (\frenchspacing{Fig. 2}, dotted lines) yields the fraction of refrigerant mass 
monitored with the present sound acquisition chain, 
$34\pm 2\%$ ($74\pm 4\%$) of the total at $P\!=$2 atm ($P\!=$1 
atm), 
a decisive datum to obtain dark matter limits.
\begin{figure}[tbp]
\epsfxsize = \hsize \epsfbox{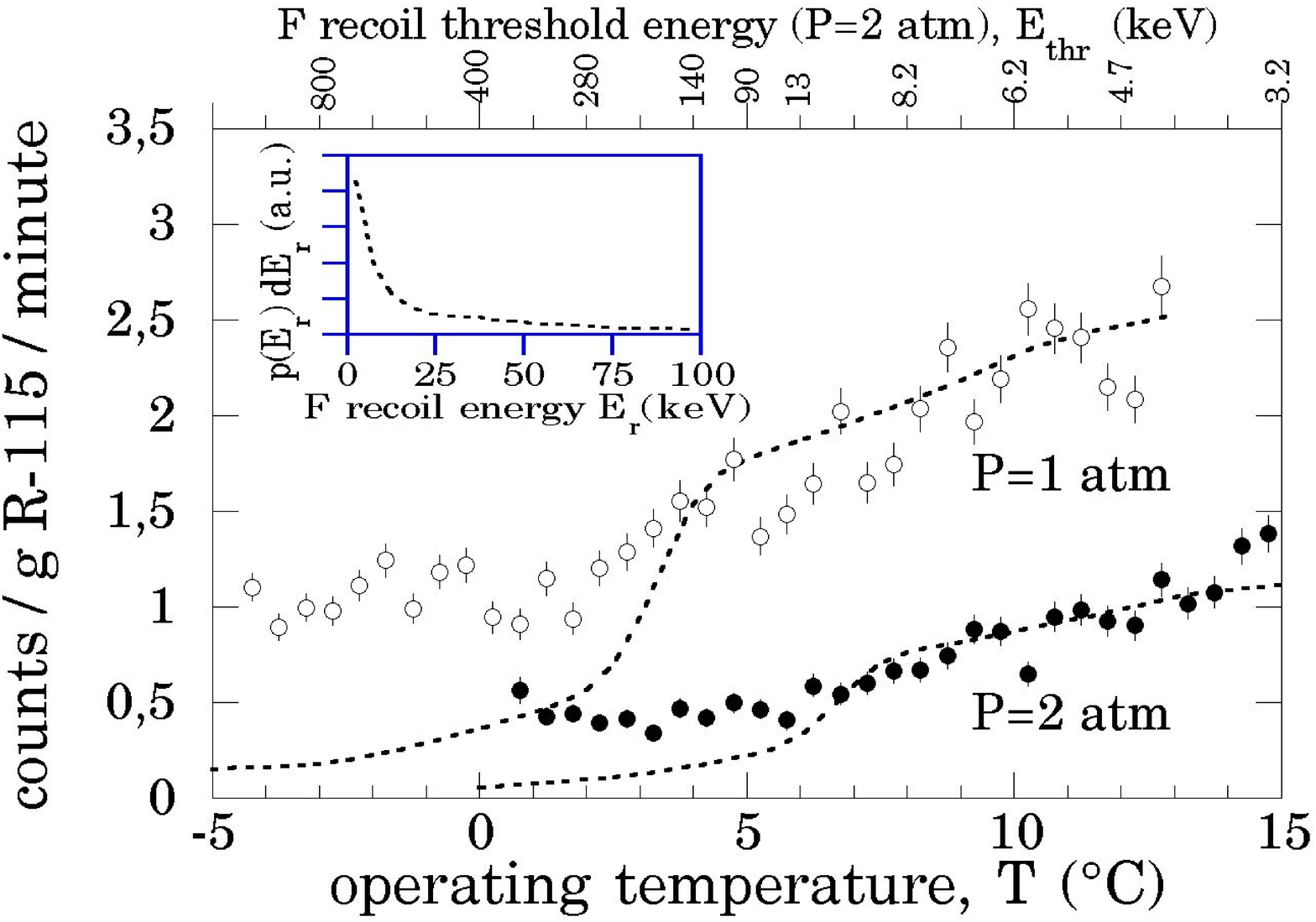}
\caption{$^{252}$Cf neutron calibration of SIMPLE modules at the 
TIS/RP bench (CERN), 
compared with Monte Carlo expectations (dotted lines, see text).  
The signal-to-noise ratio 
was $>30$ at all times. {\it Insert}: calculated energy spectrum of 
F recoils during the irradiations.} 
\end{figure}
The installation 500 m underground of modules identical in 
preparation and sound detection system to those utilized in 
$^{252}$Cf calibrations started in July 1999. A decommissioned 
nuclear missile control center has been converted into an 
underground laboratory \cite{wwwrustrel}, facilitating this and 
other initiatives. 
The characteristics of this site (microphonic silence, 
unique electromagnetic shielding \cite{wwwrustrel}) 
make it specially adequate for rare-event searches. 
Modules are placed inside a 
thermally-regulated water bath, surrounded by three layers of 
sound and thermal insulation. A 700 l water neutron moderator, 
resting on a vibration absorber, completes the shielding. Events in 
the modules and in external microphones are 
time-tagged, 
allowing to filter-out the small fraction ($\sim\!15$\%) of signals 
correlated to human activity in the immediate vicinity of the 
experiment. $P$ and $T$ are continually logged. 
The signal waveforms are digitally stored, but no event rejection based 
on pulse-shape considerations \cite{zacek2} is performed at this 
stage, eluding the criticisms \cite{gerbier} associated to some 
WIMP searches in which large data cuts are made. 

The raw counting rate from the first SIMPLE module operated in 
these conditions appears in \frenchspacing{Fig. 3}.  Accounting for 
sound 
detection efficiency and a 62\% fluorine mass fraction in R-115, 
limits can be extracted 
on the spin-dependent WIMP-proton cross section $\sigma_{Wp}$ 
(\frenchspacing{Fig. 3}). 
The cosmological parameters and method in \cite{smith} are 
used in the calculation of WIMP elastic scattering rates, which are 
then compared to the observed uncut nucleation rate at 
$T=10^{\circ}$C or $14^{\circ}$C, depending on WIMP mass. The 
expected nucleation rate 
at $T$ (i.e., integrated for recoil energies 
$>\!E_{thr}(T)$) from a candidate at the 
edge of the sensitivity of the leading DAMA experiment \cite{review} 
($\sim\!1.5\cdot10^{4}$ kg-day of NaI) 
is offered as a reference in \frenchspacing{Fig. 3}: 
SIMPLE sensitivity is presently limited by the large statistical 
uncertainty associated to a short exposure, and not yet by background 
rate.
A considerable improvement 
is expected after the ongoing expansion of the bath to accommodate 
up to 16 modules. In parallel to this, plastic module caps are being 
replaced by a sturdier design: runs using refrigerant-free modules 
show that 
{\it a majority} of the recorded events arise from pressure 
microleaks, correlated 
to the sense of $T$ ramping, 
able to stimulate the piezoelectric sensor. It must also 
be kept in mind that a $T-$independent, flat background implies a 
null WIMP signal, albeit this eventual 
approach to data analysis can only 
be exploited after a large reduction in statistical uncertainty is 
achieved. \begin{figure}[tbp]
\epsfxsize = \hsize \epsfbox{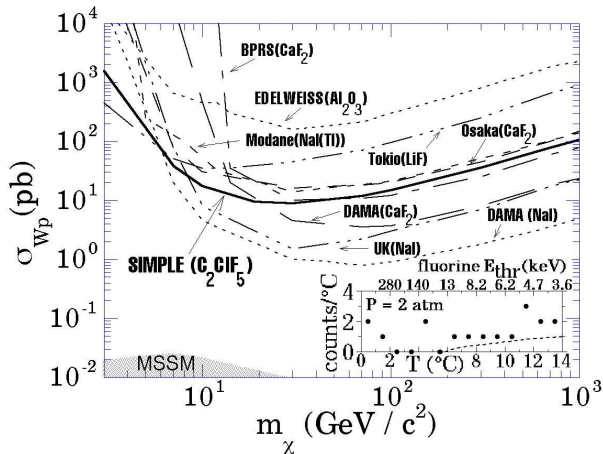}
\caption{95\% C.L. limits on $\sigma_{Wp}$ extracted from 0.19 
kg-day of SDD exposure, compared with other experiments [8]. 
``MSSM'' marks the tip of the region where a lightest 
supersymmetric partner is expected. {\it Insert}: Counting rate 
in 
the module ($9.2\pm0.1$ g R-115, 
$\Delta T=-0.75^{\circ}$C/day). The dotted line is the expected 
signal (corrected for 34\% detection efficiency) 
from a WIMP of mass $m_{\chi}=$10 GeV and $\sigma_{Wp}=$ 
5 pb: present sensitivity is still limited by low statistics.} 
\end{figure}
The importance of the spin-dependent WIMP interaction 
channel (where F is the optimal target \cite{john}) has been 
recently 
stressed by its relative insensitivity to CP-violation 
parameter values, which may otherwise severely reduce 
coherent interaction rates \cite{paolo,njop}. Nevertheless,  
$CF_{3}Br$  modules 
able to  
exploit coherent couplings are presently under development.
The intrinsic insensitivity of SDDs to most undesirable 
backgrounds, low cost of materials involved and simplicity of 
production and operation opens a new door to dark matter 
detection.

We thank the Communaut\'e des Communes du Pays d'Apt 
and French Ministry of Defense for supporting the 
conversion of the underground site. Our gratitude goes to M. 
Auguste, J. Bourges, G. Boyer, R. Brodzinski, A. Cavaillou, 
COMEX-PRO, 
M. El-Majd, M. Embid, L. Ibtiouene, 
IMEC, J. Matricon, M. Minowa, Y.H. Mori, T. Otto, 
G. Roubaud, M. Same and C.W. Thomas. 

\end{document}